\documentclass[aps,superscriptaddress,preprintnumbers,showpacs]{revtex4}
\usepackage{amssymb}
\usepackage{dcolumn}
\usepackage{bm}
\usepackage{graphicx}
\usepackage{booktabs}
\usepackage{multirow}

\newcommand{\e}{{\mathrm{e}}}
\renewcommand{\i}{{\mathrm{i}}}

\begin{document}

\newcommand*{\PKU}{School of Physics and State Key Laboratory of Nuclear Physics and
Technology, Peking University, Beijing 100871,
China}\affiliation{\PKU}
\newcommand*{\CHEP}{Center for High Energy Physics, Peking University, Beijing 100871, China}\affiliation{\CHEP}

\title{Simple parametrization of neutrino mixing matrix}

\author{Bo-Qiang Ma}\email{mabq@pku.edu.cn}\affiliation{\PKU}\affiliation{\CHEP}

\begin{abstract}
We propose simple forms of neutrino mixing matrix in analogy with
the Wolfenstein parametrization of quark mixing matrix, by adopting
the smallest mixing angle $\theta_{13}$ as a measure of expansion
parameters with the tribimaximal pattern as the base matrix. The
triminimal parametrization technique is utilized to expand the
mixing matrix under two schemes, i.e., the standard Chau-Keung (CK)
scheme and the original Kobayashi-Maskawa (KM) scheme. The new
parametrizations have their corresponding Wolfenstein-like
parametrizations of quark mixing matrix, and therefore they share
the same intriguing features of the Wolfenstein parametrization. The
newly introduced expansion parameters for neutrinos are connected to
the Wolfenstein parameters for quarks via the quark-lepton
complementarity.
\end{abstract}


\maketitle



The recent establishment of a non-zero and relatively large value of
the smallest neutrino mixing angle $\theta_{13}$ by a number of
experiments~\cite{expt,daya-bay,Ahn:2012nd}
can be considered as a signal for the high precision era of neutrino
oscillations. Among the four parameters for the description of
neutrino mixing matrix, the three mixing angles have been measured
to rather high precision, with only the CP violating phase $\delta$
being unknown yet. It is thus time to reconsider our understanding
of the neutrino mixing pattern and seek a simple parametrization of
neutrino mixing matrix, in analogy with the Wolfenstein
parametrization~\cite{wolf} of quark mixing matrix.

In the standard model of particle physics, the mixing
is well described by the Cabibbo-Kobayashi-Maskawa matrix
$V_{\mathrm{CKM}}$~\cite{CKM1,CKM2} for quarks and the
Pontecorvo-Maki-Nakagawa-Sakata~(PMNS) matrix
$U_{\mathrm{PMNS}}$~\cite{PMNS} for leptons. Among many options, the
standard parametrization, i.e., the Chau-Keung (CK)
scheme~\cite{CK}, expresses the mixing matrix by three angles
$\theta_{12}$, $\theta_{13}$, $\theta_{23}$ and one CP-violating
phase angle $\delta$ in a form
\begin{eqnarray}
V/U = \left (
\begin{array}{ccc}
c^{~}_{12} c_{13}         & s^{~}_{12} c_{13}       & s_{13}
\e^{-\i\delta}\cr -c^{~}_{12} s_{23} s_{13} \e^{i\delta}- s^{~}_{12}
c_{23} & -s^{~}_{12} s_{23} s_{13}\e^{\i\delta} + c^{~}_{12} c_{23}
& s_{23} c_{13} \cr -c^{~}_{12} c_{23} s_{13} \e^{\i\delta}+
s^{~}_{12} s_{23} & -s^{~}_{12} c_{23} s_{13} \e^{\i\delta}-
c^{~}_{12} s_{23} & c_{23} c_{13}
\end{array}
\right), \label{CK}
\end{eqnarray}
where $s_{ij}={\rm sin}\theta_{ij}$ and $c_{ij}={\rm
cos}\theta_{ij}$ ($i,j=1,2,3$). For describing the quark mixing,
another simple form of parametrization, i.e., the Wolfenstein
parametrization~\cite{wolf}, was proposed with the re-definition
that $s_{12}=\lambda$, $s_{23}=A\lambda^2$ and
$s_{13}\e^{\i\delta}=A\lambda^3(\rho+\i\eta)$. Its explicit form at
the accuracy of ${\cal O}(\lambda^4)$ is
\begin{equation}
V_{\rm CKM} = \left (
\begin{array}{ccc}
1-\frac{1}{2}\lambda^{2}    & \lambda   & A\lambda^{3}(\rho
-\i \eta) \\
-\lambda    & 1-\frac{1}{2}\lambda^{2}    &
A\lambda^{2} \\
A\lambda^{3}(1-\rho-\i \eta)  & -A\lambda^{2}     & 1
\end{array}
\right )+{\cal O}(\lambda^4) \;.
\end{equation}
Such a parametrization has a number of advantages, such as
simplicity in form, convenience for use, and hierarchical feature in
structure, thus it is widely adopted in theoretical analysis and
phenomenological applications.

The Wolfenstein parametrization can be considered as expansions in
orders of $\lambda$ around the unit matrix. For the neutrino mixing,
the situation becomes complicated as expansions around the unit
matrix is inadequate due to the large values of neutrino mixing
angles $\theta_{12}$ and $\theta_{23}$. There have been attempts to
parameterize the PMNS matrix based on some mixing patterns with
finite mixing angles, such as the  bimaximal~(BM) pattern~\cite{bi}
with $\theta_{12}=\theta_{23}=45^\circ$ and the tribimaximal~(TB)
pattern~\cite{tri} with $\theta_{12}=35.26^\circ$ and
$\theta_{23}=45^\circ$, and the base matrices of the two patterns
are
\begin{eqnarray}
U_{\rm BM}=\left(
\begin{array}{ccc}
\frac{1}{\sqrt{2}} & \frac{1}{\sqrt{2}}  & 0 \\
-\frac{1}{2} & \frac{1}{2} & \frac{1}{\sqrt{2}}  \\
\frac{1}{2} & -\frac{1}{2} & \frac{1}{\sqrt{2}}
\end{array}\right), \ \
U_{\rm TB}=\left(
\begin{array}{ccc}
\sqrt{\frac{2}{3}}& \frac{1}{\sqrt{3}} & 0 \\
-\frac{1}{\sqrt{6}} & \frac{1}{\sqrt{3}} & \frac{1}{\sqrt{2}} \\
\frac{1}{\sqrt{6}} &  -\frac{1}{\sqrt{3}}& \frac{1}{\sqrt{2}}
\end{array}\right),
\label{TBform}
\end{eqnarray}
where an additional factor $P_\nu={\rm
Diag}\{e^{-\i\alpha/2},e^{-\i\beta/2},1\}$ should be multiplied for
the Majorana neutrino case. In both of the above mixing patterns,
the mixing angle $\theta_{13}$ is chosen to be zero.

A large and non-zero $\theta_{13}$ poses a challenge to previous
mixing patterns such as the TB pattern which has received studies
using basic symmetries. We need to make the upgrade of previous
parametrizations which are mostly based on the speculation of a
small $\theta_{13}$~\cite{lisw}. A new mixing pattern has been
suggested~\cite{Zheng:2011uz} based on a self-complementary
relation~\cite{Zhang:2012xu} $\theta_{12}^{\nu} + \theta_{13}^{\nu}
=\theta_{23}^{\nu}= 45^\circ$ between neutrino mixing angles. The
new base matrix and also the consequent Wolfenstein-like
parametrization are complicated. Therefore a new strategy for the
parametrization of the PMNS matrix should be re-designed.

In this paper we take the smallest mixing angle $\theta_{13}$ as a
Wolfenstein-like parameter for the PMNS matrix expansion. We utilize
the triminimal parametrization
technique~\cite{triminimal,King:2007pr} and adopt the three
parameters $s\sim 0.1$, $a\sim 1$ and $b\sim 1$ from the following
relations
\begin{eqnarray}
&&\sin{\theta_{12}}=\frac{1}{\sqrt{3}}(1 -2 a s^3), \nonumber \\
&&\sin{\theta_{23}}=\frac{1}{\sqrt{2}}(1 -b s^2), \nonumber \\
&&\sin{\theta_{13}}=\sin0^\circ+ \sqrt{2}s=\sqrt{2} s,
\label{TriminimalP}
\end{eqnarray}
where the tri-maximal bases are taken to be the TB mixing pattern.
From Eq.~(\ref{TriminimalP}) we get the corresponding trigonometric
functions
\begin{eqnarray}
&&\cos{\theta_{12}}=\sqrt{1-\sin^2{\theta_{12}}}=\sqrt{\frac{2}{3}}\left(1+a s^3\right)+{\cal O}(s^4\to s^6), \nonumber \\
&&\cos{\theta_{23}}=\sqrt{1-\sin^2{\theta_{23}}}=\frac{1}{\sqrt{2}}\left( 1+ b s^2\right)+{\cal O}(s^4), \nonumber \\
&&\cos{\theta_{13}}=\sqrt{1-\sin^2{\theta_{13}}}=1- s^2+ {\cal
O}(s^4). \label{Triminimal}
\end{eqnarray}
Substituting the above expressions into Eq.~(\ref{CK}) we obtain
\begin{eqnarray}
U_{\rm{PMNS}}&=&\left(\begin{array}{ccc}
\sqrt{\frac{2}{3}}\left(1- s^2 +a s^3 \right)& \frac{1}{\sqrt{3}}\left(1- s^2 -2 a s^3\right) & 0  \\
-\frac{1}{\sqrt{6}}\left(1+b s^2 -2a s^3 \right) &
\frac{1}{\sqrt{3}}\left(1+ b s^2 +a s^3 \right) &
\frac{1}{\sqrt{2}}\left(1 - s^2-b  s^2  \right) \\
\frac{1}{\sqrt{6}}\left(1-b s^2 -2 a s^3 \right) &
-\frac{1}{\sqrt{3}}\left(1-b s^2 +a s^3 \right)&
\frac{1}{\sqrt{2}}\left(1 - s^2+b s^2 \right)
\end{array}\right)  \nonumber \\&&+
\left(\begin{array}{ccc}
0& 0 & \sqrt{2} s \e^{-\i\delta}  \\
-\sqrt{\frac{2}{3}} s \e^{\i\delta} \left(1-b  s^2
 \right) & -\frac{1}{\sqrt{3}} s \e^{\i\delta}
\left(1-b s^2 \right) &
0 \\
-\sqrt{\frac{2}{3}} s \e^{\i\delta}\left(1+b s^2 \right)&
-\frac{1}{\sqrt{3}} s \e^{\i\delta}\left(1+b s^2 \right)& 0
\end{array}\right)
+{\cal O}(s^4), \label{TBtriminimal}
\end{eqnarray}
which can be re-written in a form
\begin{eqnarray}
U_{\rm{PMNS}}&=& U_{\mathrm{TB}} \circledast
\left(\begin{array}{ccc}
1- s^2 +a s^3 & 1- s^2  -2a s^3 & 0  \\
1+b s^2 -2a s^3 & 1+b s^2 +a s^3 &
1 - s^2-b s^2  \\
1-b s^2 -2a s^3  & 1-b s^2 +a s^3 & 1 - s^2+b s^2
\end{array}\right) \nonumber \\ &&
-\frac{1}{\sqrt{3}} s \e^{\i\delta}
 \left(\begin{array}{ccc}
0 & 0 & -\sqrt{6}  \e^{-2 \i\delta}  \\
\sqrt{2}\left(1-b s^2 \right) & \left(1-b s^2 \right) &
0 \\
\sqrt{2}\left(1+b s^2 \right) & \left(1+b s^2 \right) & 0
\end{array}\right)+{\cal O}(s^4)
, \label{TBtriminimal2}
\end{eqnarray}
where the sign $\circledast$ means direct multiplications between
the corresponding elements of the two matrices. For $a=b=0$, the
parametrization reduces to the latest parametrization by
King~\cite{King} with a similar form
\begin{eqnarray}
U_{\rm{PMNS}}&=&\left(\begin{array}{ccc}
\sqrt{\frac{2}{3}}\left(1- s^2 \right)& \frac{1}{\sqrt{3}}\left(1- s^2\right) & \sqrt{2}s \e^{-\i\delta}  \\
-\frac{1}{\sqrt{6}}-\sqrt{\frac{2}{3}} s \e^{\i\delta} &
\frac{1}{\sqrt{3}}-\sqrt{\frac{1}{3}} s \e^{\i\delta} &
\frac{1}{\sqrt{2}}\left(1 - s^2 \right) \\
\frac{1}{\sqrt{6}} -\sqrt{\frac{2}{3}} s \e^{\i\delta}&
-\frac{1}{\sqrt{3}}-\sqrt{\frac{1}{3}} s \e^{\i\delta}&
\frac{1}{\sqrt{2}}\left(1 - s^2  \right)
\end{array}\right)  +{\cal O}(s^2),
\label{TBtriminimal3}
\end{eqnarray}
from which we see that the deviation could be of the order ${\cal
O}(s^2)$ instead of ${\cal O}(s^3)$ if we adopt $\theta_{23} \approx
40^\circ$ as our input for $b \sim 1$.

Eq.~(\ref{TBtriminimal2}) can be considered as our proposal for a
simple Wolfenstein-like parametrization for neutrino mixing with
respect to the standard parametrization Eq.~(\ref{CK}). It is a full
parametrization with four independent parameters $s$, $a$, $b$ and
$\delta$, corresponding to the three Euler angles $\theta_{13}$,
$\theta_{12}$, $\theta_{23}$ and the CP violating phase $\delta$ in
the CK-scheme. One advantage of Eq.~(\ref{TBtriminimal2}) is that
the leading order of the matrix corresponds to the tribimaximal
pattern $U_{\mathrm{TB}}$. In similar to the Wolfenstein
parametrization in which the largest mixing angle
$\theta_{\mathrm{C}}$ serves as the main expansion parameter
$\lambda=\sin(\theta_{\mathrm{C}})$, the main expansion parameter
$s$ for neutrinos corresponds to the smallest mixing angle
$\theta_{13}$ by $\sqrt{2}s=\sin(\theta_{13})$. Another advantage of
Eq.~(\ref{TBtriminimal2}) is that the CP violating terms are
expressed in the matrix with the factor $\e^{\i\delta}$, therefore
this parametrization is also convenient for the analysis of CP
violation in neutrino oscillations. Taking the latest results from
phenomenological analysis~\cite{Zhang:2012bk} as our input
\begin{eqnarray}
\theta_{12}&=&(33.96^{+1.03}_{-0.99}(^{+3.22}_{-2.91}))^\circ;\nonumber\\
\theta_{23}&=&(40.40^{+4.64}_{-1.74}(^{+12.77}_{-4.64}))^\circ;\nonumber\\
\theta_{13}&=&(9.07\pm0.63(\pm1.89))^\circ,
\end{eqnarray}
we obtain $s\approx 0.111$, $a \approx 11.7$ and $b \approx  6.71$.

In alternative to the CK-scheme, the Kobayashi-Maskawa (KM)
scheme~\cite{CKM2} 
is also been shown to own some intriguing features which are
convenient for
applications~\cite{koide,Koide:2008yu,boomerang,Li:2010ae,qinnan,Ahn:2011it}.
One example is a simple Wolfenstein-like parametrization of the
form~\cite{qinnan}
\begin{eqnarray}
V=\left(\begin{array}{ccc}
1-\frac{\lambda^2}{2}&\lambda&\e^{-\i \tilde{\delta}}h\lambda^3\\
-\lambda&1-\frac{\lambda^2}{2}&(f+\e^{-\i \tilde{\delta}}h)\lambda^2\\
f\lambda^3&-(f+\e^{\i\tilde{\delta}}h)\lambda^2&1
\end{array}\right)
+\mathcal {O}(\lambda^4)\;,\label{newwolf}
\end{eqnarray}
with respect to the re-phased form~\cite{qinnan} of mixing matrix in
the KM-scheme
\begin{eqnarray} V_{\rm
KM}&=&\left(
\begin{array}{ccc}
c_1     & s_1c_3                     & s_1s_3 \e^{-\i\tilde{\delta} }         \\
-s_1c_2  & c_1c_2c_3-s_2s_3\e^{\i \tilde{\delta}} & s_2c_3 + c_1c_2s_3 \e^{-\i\tilde{\delta}} \\
s_1s_2  & -c_1s_2c_3-c_2s_3\e^{\i \tilde{\delta}} &
c_2c_3-c_1s_2s_3\e^{-\i \tilde{\delta}}
\end{array}
\right)\;,\label{KM}
\end{eqnarray}
in which $s_i=\sin\theta_i$ and $c_i=\cos\theta_i$ correspond to
Euler angles $\theta_i$ $(i=1,2,3)$, and $\tilde{\delta}$ is the
CP-violating phase in the KM parametrization. One advantage of the
KM-scheme is that it allows for almost a perfect maximal CP
violation of the quark
mixing~\cite{koide,Koide:2008yu,boomerang,Li:2010ae,qinnan,Ahn:2011it},
i.e., $\tilde{\delta}^{\mathrm{quark}}=90^\circ$. With an Ansatz of
maximal CP violation $\tilde{\delta}^{\mathrm{lepton}}=90^\circ$ for
also leptons~\cite{Zhang:2012ys}, we get the mixing angles in the KM
parametrization~\cite{Zhang:2012bk}
\begin{eqnarray}
\theta_1&=&(35.01^{+1.01}_{-0.97})^\circ;\nonumber\\
\theta_2&=&(39.86^{+5.14}_{-1.97})^\circ;\nonumber\\
\theta_3&=&(15.96^{+1.11}_{-1.18})^\circ.
\end{eqnarray}
We introduce the three parameters
\begin{eqnarray}
&&\sin{\theta_{1}}=\frac{1}{\sqrt{3}}(1 -2 \tilde{a} \tilde{s}^4), \nonumber \\
&&\sin{\theta_{2}}=\frac{1}{\sqrt{2}}(1  -\tilde{b} \tilde{s}^2), \nonumber \\
&&\sin{\theta_{3}}=\sqrt{2} \tilde{s}, \label{TriminimalKM}
\end{eqnarray}
from which we get
\begin{eqnarray}
&&\cos{\theta_{1}}=\sqrt{1-\sin^2{\theta_{1}}}=\sqrt{\frac{2}{3}}\left(1+ \tilde{a}  \tilde{s}^4\right)+{\cal O}(\tilde{s}^8), \nonumber \\
&&\cos{\theta_{2}}=\sqrt{1-\sin^2{\theta_{2}}}=\frac{1}{\sqrt{2}}\left( 1+ \tilde{b} \tilde{s}^2\right)+{\cal O}(\tilde{s}^4), \nonumber \\
&&\cos{\theta_{3}}=\sqrt{1-\sin^2{\theta_{3}}}=1- \tilde{s}^2+ {\cal
O}(\tilde{s}^4), \label{TriminimalPP}
\end{eqnarray}
where $\tilde{s}\approx 0.194$, $\tilde{a} \approx 2.20$ and
$\tilde{b}\approx 2.48$. Substituting the above trigonometric
functions into Eq.~(\ref{KM}), we obtain the simple parametrization
of the PMNS matrix with respect to the KM-scheme
\begin{eqnarray}
U_{\rm{PMNS}}&=& U_{\mathrm{TB}} \circledast
\left(\begin{array}{ccc}
1 & 1-\tilde{s}^2   & 0  \\
1+ \tilde{b} \tilde{s}^2  & 1- \tilde{s}^2+ \tilde{b} \tilde{s}^2 &
1 - \tilde{s}^2- \tilde{b} \tilde{s}^2  \\
1- \tilde{b} \tilde{s}^2   & 1-\tilde{s}^2- \tilde{b} \tilde{s}^2
 & 1 - \tilde{s}^2+ \tilde{b} \tilde{s}^2
\end{array}\right) 
+ \tilde{s} \e^{\i\tilde{\delta}}
 \left(\begin{array}{ccc}
0 & 0 & \sqrt{{2}/{3}}  \e^{-2 \i\tilde{\delta}}  \\
0& -1 & \sqrt{{2}/{3}} \e^{-2 \i\tilde{\delta}}  \\
0& -1 & -\sqrt{{2}/{3}} \e^{-2 \i\tilde{\delta}}
\end{array}\right)+{\cal O}(\tilde{s}^3). \label{TBtriminimal2KM}
\end{eqnarray}
In comparison with Eq.~(\ref{TBtriminimal2}), this form of the PMNS
matrix looks more simple.

We now provide connections between the newly introduced parameters
of neutrinos and the Wolfenstein parameters of quarks via the
quark-lepton complementarity
(QLC)~\cite{QLC1,phenomenology,Zhang:2012zh}. The QLC in forms of
mixing matrices can lead to a simple relation $U_{e3}\approx
\lambda/\sqrt{2}$~\cite{Zheng:2011uz,Li:2005yj,Qin:2011bq}, thus we
have
\begin{equation}
s\approx{\lambda}/{2}, \ \ \tilde{s}\approx{{\sqrt{3}\lambda}/{2}} ,
\end{equation}
where the first relation can also be obtained from alternative
arguments~\cite{King,Antusch}. From the QLC in the form of mixing
angles, i.e., $\theta^q_{23}+\theta^l_{23}=45^{\circ}$,
we arrive at the relation
$b s^2\sim  \tilde{b} \tilde{s}^2 \sim \kappa A \lambda^2$, with
$\kappa_{23} =(45^\circ-\theta^l_{23})/\theta^q_{23}=1.96$ and
$\kappa_{2} =(45^\circ-\theta^l_{2})/\theta^q_{2}=2.18$ being the
adjusting factors of the data.
Then we get
\begin{eqnarray}
b \sim
4\kappa_{23} A , \ \ \tilde{b} \sim
4\kappa_{2}A/3.
\end{eqnarray}
Thus our newly proposed simple parametrizations
Eqs.~(\ref{TBtriminimal2}) and (\ref{TBtriminimal2KM}) can be also
considered as expansions in terms of the Wolfenstein parameters
$\lambda$ and $A$ of quarks. Adopting $\lambda=0.2253$ and $A=0.808$
we obtain
\begin{eqnarray}
s\approx 0.113,~~ b\sim 6.32; \ \  \tilde{s}\approx0.195,
~~\tilde{b}\sim 2.37,
\end{eqnarray}
which are compatible with the above estimated values from neutrino
oscillation data. Further accurate measurements can test and improve
the above suggested connections between quarks and leptons.

There is uncertainty with the choice of the powers of $s$ or
$\tilde{s}$ with respect to $a$, $b$ or $\tilde{a}$, $\tilde{b}$.
For example, we can alternatively choose
$\sin{\theta_{12}}=\frac{1}{\sqrt{3}}(1 -2 a s^2)$ in the CK-Scheme
or $\sin{\theta_{1}}=\frac{1}{\sqrt{3}}(1 -2 \tilde{a} \tilde{s}^3)$
in the KM-Scheme, so that the new PMNS matrix is
\begin{eqnarray}
U_{\rm{PMNS}}&=& U_{\mathrm{TB}} \circledast
\left(\begin{array}{ccc}
1- s^2 +a s^2 & 1- s^2  -2a s^2 & 0  \\
1+b s^2 -2a s^2 & 1+b s^2 +a s^2 &
1 - s^2-b s^2  \\
1-b s^2 -2a s^2  & 1-b s^2 +a s^2 & 1 - s^2+b s^2
\end{array}\right) \nonumber \\ &&
-\frac{1}{\sqrt{3}} s \e^{\i\delta}
 \left(\begin{array}{ccc}
0 & 0 & -\sqrt{6}  \e^{-2 \i\delta}  \\
\sqrt{2}\left(1-b s^2 \right) & \left(1-b s^2 \right) &
0 \\
\sqrt{2}\left(1+b s^2 \right) & \left(1+b s^2 \right) & 0
\end{array}\right)+{\cal O}(s^4)
, \label{TBtriminimal2b}
\end{eqnarray}
with $s\approx 0.111$, $a \approx 1.31$ and $b \approx  6.71$, or
\begin{eqnarray}
U_{\rm{PMNS}}&=& U_{\mathrm{TB}} \circledast
\left(\begin{array}{ccc}
1+ \tilde{a} \tilde{s}^3 & 1-\tilde{s}^2 -2 \tilde{a} \tilde{s}^3  & 0  \\
1+ \tilde{b} \tilde{s}^2 -2 \tilde{a} \tilde{s}^3  & 1- \tilde{s}^2+
\tilde{b} \tilde{s}^2 + \tilde{a} \tilde{s}^3&
1 - \tilde{s}^2- \tilde{b} \tilde{s}^2  \\
1- \tilde{b} \tilde{s}^2 -2 \tilde{a} \tilde{s}^3   & 1-\tilde{s}^2-
\tilde{b} \tilde{s}^2 + \tilde{a} \tilde{s}^3
 & 1 - \tilde{s}^2+ \tilde{b} \tilde{s}^2
\end{array}\right) \nonumber \\ &&
+ \frac{1}{\sqrt{3}} \tilde{s} \e^{\i\tilde{\delta}}
 \left(\begin{array}{ccc}
0 & 0 & \sqrt{{2}}  \e^{-2 \i\tilde{\delta}}  \\
0& -\sqrt{3}(1- \tilde{b} \tilde{s}^2) & \sqrt{{2}}(1+ \tilde{b} \tilde{s}^2) \e^{-2 \i\tilde{\delta}}  \\
0& -\sqrt{3}(1+ \tilde{b} \tilde{s}^2) & -\sqrt{{2}} (1- \tilde{b}
\tilde{s}^2) \e^{-2 \i\tilde{\delta}}
\end{array}\right)+{\cal O}(\tilde{s}^4), \label{TBtriminimal2KMb}
\end{eqnarray}
with $\tilde{s}\approx 0.194$, $\tilde{a} \approx 0.428$ and
$\tilde{b}\approx 2.48$. The corresponding powers of $s$ or
$\tilde{s}$ related to $b$ or $\tilde{b}$ could be $3$ or higher
instead of $2$, if the experimental value for $\theta_{23}$ or
$\theta_{2}$ is more close to $45^\circ$. Further improved
measurements of neutrino mixing data or theoretical progress on the
underlying quark-lepton connections can help to determine the
explicit powers.

In conclusion, we suggest simple forms of the PMNS matrix with some
intriguing features possessed by the Wolfenstein parametrization of
quark mixing matrix, such as full form of mixing matrix with four
independent parameters, simplicity in form for convenient
applications especially for CP violation study, and also
hierarchical structure as expansions around the tribimaximal
pattern. Though the explicit forms of parametrization, such as the
powers of $s$ or $\tilde{s}$ with respect to the parameters $a$, $b$
or $\tilde{a}$, $\tilde{b}$, might change according to future
improved measurements of neutrino mixing parameters,
parametrizations in the same spirit of the Wolfenstein
parametrization might have more chance for wide applications.

\begin{acknowledgments}
 This work is partially supported by National Natural
Science Foundation of China (Grants No.~11021092, No.~10975003,
No.~11035003, and No.~11120101004).
\end{acknowledgments}

\end{document}